# Radiation Induced Point and Cluster-Related Defects with Strong Impact to Damage Properties of Silicon Detectors

Ioana Pintilie[*,a], Gunnar Lindström[b], Alexandra Junkes[b], Eckhart Fretwurst[b]

[a]*Institute of Materials Physics NIMP, Str. Atomistilor 105 bis, RO-77125 Bucharest, Romania*
[b]*Institute for Experimental Physics, University of Hamburg, Luruper Chaussee 149, 22761 Hamburg, Germany*

**Abstract**

This work focuses on the investigation of radiation induced defects responsible for the degradation of silicon detectors. Comparative studies of the defects induced by irradiation with $^{60}$Co-$\gamma$ rays, 6 and 15 MeV electrons, 23 GeV protons and reactor neutrons revealed the existence of point defects and cluster related centers having a strong impact on damage properties of Si diodes. The detailed relation between the "microscopic" reasons as based on defect analysis and their "macroscopic" consequences for detector performance are presented. In particular, it is shown that the changes in the Si device properties after exposure to high levels of $^{60}$Co-$\gamma$ doses can be completely understood by the formation of two point defects, both depending strongly on the oxygen concentration in the silicon bulk. Specific for hadron irradiation are the annealing effects which decrease respectively increase the originally observed damage effects as seen by the changes of the depletion voltage. A group of three cluster related defects, revealed as deep hole traps, proved to be responsible specifically for the reverse annealing. Their formation is not affected by the oxygen content or silicon growth procedure suggesting that they are complexes of multi-vacancies located inside extended disordered regions.

*Key words:* Silicon detectors, EPI, FZ, MCz, defect analysis, TSC measurements, $N_{eff}$ predictions, S-LHC tracker

## 1. Introduction

One of the most challenging applications for silicon detectors is given by their use in the inner tracking region of forthcoming colliding beam experiments as e.g. LHC (Large Hadron Collider at the European research center CERN) and especially its planned upgrade (S-LHC), the International Linear Collider (ILC) or high brilliance photon sources like the European XFEL foreseen also for the next decade [1–3]. Segmented silicon sensors (micro-strip and pixel devices) are at present the most precise electronic tracking detectors in high energy physics experiments (HEP). This paper addresses the understanding of radiation damage effects resulting from the non-ionizing energy loss (NIEL) leading to displacement damage in the silicon bulk. Surface and interface related effects caused by ionization are on the other hand of considerable importance only for applications in environments with high X-ray doses (as e.g. in the European XFEL), a topic not addressed here.

For the S-LHC the detectors closest to the beam have to perform at hadron fluences up to several $10^{16}$ cm$^{-2}$ under complex, long term operation scenarios [4–6]. The limitations for their practical application in the hadron colliders are caused by irradiation induced defects leading to changes in the effective doping concentration ($N_{eff}$) respectively full depletion voltage ($V_{dep}$), the reverse current at the depletion voltage ($I_{dep}$) and the degradation in the charge collection efficiency (CCE) [5–14]. These device properties are subject to changes not only during irradiation but also during beam-off periods. Especially the long term annealing effects in $N_{eff}$ increase the initial depletion voltage and therefore they are of extreme importance envisioning extended operational periods of several years. In order to avoid a non-tolerable increase of the depletion voltage, the detectors have to be cooled not only during operation but also for the beam-off periods throughout the entire lifetime of the experiments. It would be of considerable importance if this cold storage could be avoided. One encouraging result for improving the radiation tolerance was obtained by the CERN-RD48 Collaboration by performing an oxygen enrichment of float-zone (FZ) wafers used for detector fabrication. This defect engineering attempt was motivated by the hypothetical assumption that a large O-concentration would inhibit the formation of the $V_2O$-defect thought to be the main reason for the observed change in the effective doping. A cost effective realization of oxygen enrichment up to several $10^{17}$ cm$^{-3}$ was achieved by an in-diffusion of oxygen from the $Si-SiO_2$ interface (during treatments at high temperatures) after wafer oxidation. This procedure is called the DOFZ process (Diffusion Oxygenated Float Zone) and its main benefit is a considerable reduction of damage effects after gamma and charged hadron irradiation [15–18]. In contrast to charged hadron and gamma irradiation, neutron damage seems to be less dependent on the O-concentration. Although the obtained increase of the radiation tolerance may be sufficient to meet the requirements for LHC, the DOFZ process cannot be a solution for the much more demanding S-LHC application because of the intolerable increase of the depletion voltage up to more than 1000 V and the decrease of the CCE below the nec-essary threshold for the readout electronics. Therefore, further efforts for proper defect (and device) engineering leading to a radiation tolerance above the present level are indispensable.

[*]Corresponding author: phone: +40 21 3690170
*Email address:* ioana@infim.ro (Ioana Pintilie)

Any promising attempt for radiation hardening of the material as well as improvements by modifying the detector processing will rely on a thorough knowledge of the generation of electrically active defects which are responsible for the observed changes in the device properties at their operating temperature. This goal is addressed in our work focusing on a detailed investigation of specific radiation damage induced defects (point- and cluster-related) and their direct correlation with device performance parameters. Pion induced damage, dominating in the innermost layers of the tracking area, will result from both isolated point defects (mainly due to Coulomb interactions) and densely packed displacement regions (clusters) caused by energetic recoils from hadron reactions. A separation of both components can be best undertaken by studying damage caused by $\gamma$-irradiation (point defects) compared to neutron induced damage (clustered displacements), while the mixture of both as envisioned in pion damage can be represented by studying effects after high energetic proton irradiation. In addition low several MeV electron irradiations proved to be useful for understanding the bridge between pure point defect related and cluster dominated defects.

## 2. Experimental details and techniques

Several kinds of n-type silicon material, presently discussed as candidates foreseen for use at S-LHC, have been investigated for this purpose:

*Float-zone silicon FZ*, produced by Wacker Siltronic [19], orientation < 111 >, 300 $\mu$m thick, resistivity 3-4 k$\Omega$cm, effective doping concentration $N_d \sim 10^{12}$ cm$^{-3}$. The standard processed p$^+$-n silicon diodes (labeled as STFZ) have an oxygen content of [O] < $10^{16}$ cm$^{-3}$. O-enriched diodes (labeled as DOFZ) were obtained using an in-diffusion of oxygen from the $Si - SiO_2$ interface after wafer oxidation. The achieved O-concentration after a 72 h treatment at 1150 °C is [O] $\sim 10^{17}$ cm$^{-3}$ while the carbon concentration is at or below the detection limit (Fig. 1a).

*Magnetic Czochralski silicon MCz*, produced by Okmetic [20], orientation < 100 >, 280 $\mu$m thick, resistivity 870 $\Omega$cm, $N_d = 4.9 \times 10^{12}$ cm$^{-3}$. Due to the Czochralski process, this material has a high and almost homogeneous O-concentration around [O] $\sim 6 \times 10^{17}$ cm$^{-3}$ (see Fig. 1a).

*Thin epitaxial silicon layers EPI*, produced by ITME [21], orientation < 111 >, 72 $\mu$m thick, resistivity 170 $\Omega$cm, grown on 300 $\mu$m thick highly Sb doped Cz substrates. The effective doping concentration in EPI-diodes is $N_d \sim 2.5 \times 10^{13}$ cm$^{-3}$. In the present study both standard processed diodes (EPI-ST) as well as improved oxygen enriched ones (EPI-DO) were used. For EPI-ST the oxygen is only partly out-diffusing from the O-rich Cz substrate during the epitaxial growth, thus leading to a quite inhomogeneous distribution (Figure 1b). Measurements for EPI-ST, with 25, 50 and 75 $\mu$m thickness had been performed earlier, showing interesting effects on the average O-concentration [22]. For EPI-DO an additional diffusion step at 1100 °C for 24 h after wafer oxidation as part of the diode processing was included, leading to a homogeneous O-distribution of $6 \times 10^{17}$ cm$^{-3}$, i.e. the same value as measured in the Cz substrate (Fig. 1b). The appreciable drop of [O] in about the first 10 $\mu$m is due to out-diffusion through the front electrode as also present in the MCz sample (Fig. 1a).

In all cases diode manufacturing had been performed by the same company CiS [23], thus avoiding effects due to different process techniques. All diodes have a p$^+$ electrode of 25 mm$^2$ surrounded by a p$^+$ guard ring. The n$^+$ electrode area of 1 cm$^2$ is given by the geometrical dimension of the device. Oxygen and carbon concentration profiles in the diodes were measured by Secondary Ion Mass Spectroscopy (SIMS) at ITE [24] and are displayed in Figs. 1a and 1b. It should also be noted that for the EPI diodes depth profiles of the resistivity had been performed using Spreading Resistance measurements (SR) on bevelled samples at ITME revealing a very homogeneous distribution throughout the epitaxial-layer and an abrupt change at the EPI-substrate interface [25] (Fig. 1c).

To distinguish between point and cluster related defects, damage effects have been investigated after irradiation with $^{60}$Co-$\gamma$ rays, producing point defects only, reactor neutrons for the study of predominant cluster damage and 23 GeV protons causing a mixture of clusters and isolated point defects. The gamma irradiations were done at the BNL $^{60}$Co-$\gamma$ source with dose values up to 5 MGy [26], 6 and 15 MeV electrons were available at Stockholm [27], the TRIGA reactor in Ljubljana was used for neutron irradiations [28] and the CERN-PS served for 23 GeV proton irradiation [29].

The neutron and 23 GeV proton fluences presented in this work are 1 MeV neutron equivalent values [18], while for the electron irradiations the accumulated particle fluences are given. Further irradiation parameters which might have an impact on the results of the microscopic measurements and the macroscopic detector properties are the temperature during irradiation and the exposure time. The irradiation temperature was about 20 °C for the neutron irradiation and slightly above this value for all other exposures. On the other hand, the duration of exposure was quite different for the $\gamma$-ray and the particle irradiations. For the $\gamma$-irradiation the exposure time varied between 83 hours and 35 days for the investigated dose range of 500 kGy to 5 MGy. No annealing effects had been observed which indicates that the radiation induced point defects are stable at room temperature. The exposure times for the different particle irradiations were much shorter, i.e. in the order of 10 to 60 minutes. For these cases the situation is different because part of the damage leads to several cluster related defects which are not stable at room temperature. But the so-called short term annealing time constants of some cluster related defects, which have an impact on the reverse current (see section 5), are in the order of 30 days at room temperature. Such time constants are long compared to the irradiation time and, therefore, will not lead to a considerable self-annealing effect during irradiation. Finally it should be mentioned, that all samples were stored at temperatures below -20 °C after irradiation until the first measurements were performed.

The dependence of the defect generation on the radiation type is based on the following facts. The displacement energy in the silicon lattice, i.e. the minimum energy necessary to create a Frenkel pair (vacancy plus separated Si interstitial) is ∼ 25 eV. The majority of effects expected after $^{60}$Co-$\gamma$ irradiation is due to Compton-electrons leading to a broad distribution of recoil energies for the primary knock on Si-atom (PKA) up to 140 eV, i.e. below the threshold energy for cluster generation, considered to be 5 keV [30]. Taking recombination effects for close pairs into account, it is therefore very unlikely to directly create larger complexes than a double vacancy. In contrast to this, Si-displacements by neutron interaction are dominated by head-on collisions and e.g. for 1 MeV neutrons the mean energy transfer to the Si atom is ∼ 50 keV (note that the NIEL weighted mean energy of the Ljubljana reactor neutron spectrum is 1.7 MeV). The track length of a 50 keV PKA is about 100 nm with an extremely high density of displacements (cluster formation) [31]. On the other hand energy loss of charged hadrons is widely governed by Coulomb interactions with a PKA spectrum extending from dominant low energies (point defects) to quite large ones with a high probability of producing defect clusters [18, 32]. Finally in the case of electron irradiation the maximum PKA energy for 6 MeV electrons is 18 keV (700 displacements!) and for 15 MeV it is 250 keV and hence even the average energy transfer would already approach the value encountered for 1 MeV neutrons (see above).

The "macroscopic" device performance of the investigated diodes was measured by means of capacitance-voltage (C-V) and current-voltage (I-V) diode characteristics. The radiation induced changes in the effective doping concentration ($N_{eff}$) respectively full depletion voltage ($V_{dep}$) were determined from C-V measurements performed using a frequency of 10 kHz and with guard ring grounded. The most sensitive technique for analysis of electrically active defects is the Capacitance - Deep Level Transient Spectroscopy (DLTS). However, this method can be applied only for investigation of defects with low concentration, thus limiting the irradiation fluence to less than $10^{12}$ cm$^{-2}$ respectively the $\gamma$-dose below a few kGy (depending on the doping concentration of the material). For this situation, apart from the damage related current increase no changes in the detector performance are to be seen and thus, the defect levels responsible for the deterioration after high irradiation levels cannot be identified and correlated with the "macroscopic" characteristics of the detectors ($N_{eff}$, CCE). On the other hand the reverse current increase is much more sensitive already at small fluence levels and thus cluster related effects as expected to be relevant for the damage induced generation current as measured at depletion voltage ($I(V_{dep})$), may be very well studied in parallel with DLTS measurements.

Therefore, for larger irradiation fluences (between $10^{12}$ cm$^{-2}$ and $10^{15}$ cm$^{-2}$) we have used the Thermally Stimulated Current method (TSC) [33, 34]. It detects the centers which trap free carriers in the material and allows to determine the trapping parameters of defect levels needed to calculate their impact on the electrical properties of the device: activation energy, capture cross sections for electrons and holes, defect concentration. The TSC experimental procedure consists in cooling the sample down to a low temperature where filling of the traps is performed (by illumination or forward biasing of the diode). The TSC current measurements are then performed during heating up (with a constant heating rate of 11 K/min) under reverse bias applied to the sample. If the reverse bias is high enough to maintain full depletion of the diode during the temperature scan, the active volume of the sample is well known (defined by the guard ring controlled electrode area) and thus the defect concentrations can be calculated from the total charge released from the defects during the heating (peaks showing up in the TSC spectra).

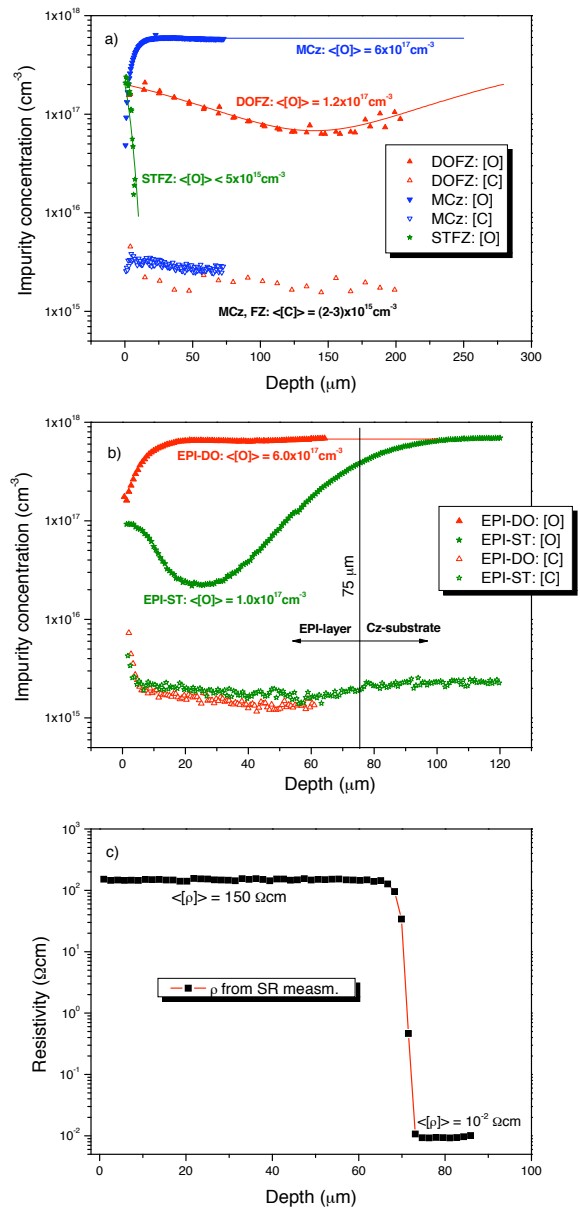

Figure 1: a) depth profiles of oxygen and carbon concentrations measured with SIMS in MCz, STFZ and DOFZ diodes; b) as a) but in EPI-ST and EPI-DO diodes; c) resistivity depth profile in a 75 $\mu$m thick EPI diode resulting from spreading resistance measurements.

## 3. Defect properties and detector performance

Many electrically active defects, induced by irradiation, are detected by DLTS and TSC experiments. Most of them ($VO$, $V_2$, $C_i$, $C_iO_i$, $C_iC_s$, $IO_2$) were already investigated in detail and no correlation with the "macroscopic" behavior of the diodes could be established [35–46]. The main characteristics of defects, from the electrical point of view, are the emission rates of carriers in the conduction and valence bands given by:

$$e_{n,p}(T) = c_{n,p}(T) \cdot N_{C,V}(T) \cdot \exp\left(\pm \frac{E_T(T) - E_{C,V}}{k_B T}\right) \quad (1)$$

$$\text{with} \quad c_{n,p}(T) = \sigma_{n,p}(T) \cdot v_{th,n,p}(T)$$

where $N_{C,V}$ = effective density of states in the conduction/valence band, $E_{C,V}$ = band edge energies, $E_T(T)$ = energy of the defect level, $\sigma_{n,p}$ = electron/hole capture cross section and $v_{th,n,p}$ is the average thermal velocity of electrons/holes.

The important features of a defect are: the capture cross sections $\sigma_{n,p}$, the defect level position in the band-gap $E_T$, the defect concentration $N_T$ and the type of the defect (acceptor or donor like). Once these characteristics are known, the influence of a defect on the space charge density as well as on the reverse current can be calculated. The contribution to $N_{eff}$ is given by the steady state occupancy of the defect levels $n_T$ that can be calculated according to the following relations resulting from the Schockley-Read-Hall statistics [34, 47]:

$$n_T^{acceptor}(T) = N_T \cdot \frac{c_n \cdot n + e_p}{e_n(T) + e_p(T) + c_n(T) \cdot n + c_p(T) \cdot p} \quad (2)$$

$$n_T^{donor}(T) = N_T \cdot \frac{c_p \cdot p + e_n}{e_n(T) + e_p(T) + c_n(T) \cdot n + c_p(T) \cdot p}$$

$$N_{eff} = \sum n_T^{donor} - \sum n_T^{acceptor}$$

where $n$, $p$ are the concentration of free electrons respectively holes in the space charge region (SCR) which can be neglected in diodes with low reverse currents. The defect contribution to the reverse current at full depletion $I(V_{dep})$ is given by [47, 48]:

$$I(V_{dep})(T) = q_0 \cdot A \cdot d \cdot \left(\sum e_n(T) \cdot n_T^{acceptor}(T) \right. \quad (3)$$
$$\left. + \sum e_p(T) \cdot n_T^{donor}(T)\right)$$

where $q_0$ is the elementary charge, $A$ and $d$ are the area and the thickness of the diode, respectively. The donors in the upper part of the gap are traps for electrons, show the Poole-Frenkel effect (a certain dependence of the emission rate of electrons on the electric field) [49] and contribute with positive space charge to $N_{eff}$ at room temperature (RT). The donors in the lower part of the gap are traps for holes, show no Poole-Frenkel effect and do not contribute to $N_{eff}$ at RT unless they are close to midgap. In contrast, acceptors in the upper part of the gap, although they are traps for electrons, do not show the Poole-Frenkel effect, and can contribute to $N_{eff}$ with negative space charge at RT only if they have close to mid-gap levels. The acceptors in the lower part of the gap are traps for holes, show the Poole-Frenkel effect and contribute with negative space charge to $N_{eff}$ at RT. The most effective centers in generating reverse current are those having close to mid-gap levels, independent of their type (acceptor or donor like).

## 4. Point defects, predominantly after $\gamma$-irradiation

Knowing the already huge efforts spent on investigating defects after hadron irradiation, with no real success in finding the defects responsible for the detector deterioration, we have decided to start our investigations with the simplest case, i.e. in situations where only point defects are generated. This was achieved by performing irradiation with $^{60}$Co- $\gamma$ rays. The diodes used for these investigations were processed on high resistivity silicon oxygen lean (STFZ) and oxygenated (DOFZ) material (see section 2.). A very pronounced beneficial effect of the oxygen content on both the current and the depletion voltage was observed [8]. While the STFZ samples undergo a space charge sign inversion - SCSI (changing the space charge sign from initially positive in n-type silicon to negative, i.e. effectively p-type material) the oxygenated diodes did not show this effect but in contrast, a gain in the positive space charge is observed with increasing irradiation dose. These "macroscopic" characteristics suggest the presence of both, a deep acceptor like defect, generated mainly in oxygen lean material, responsible for the SCSI effect and the high reverse current in STFZ material and a shallow donor in the upper part of the gap, generated mainly in oxygen rich material which causes the gain in the positive space charge with increasing irradiation dose. Detailed studies of electrically active defects by means of the TSC technique revealed indeed the generation of such point defects in the two types of materials and examples of the many experimental results are given in the following.

### 4.1. Deep acceptor $I_p$

The $I_p$-center is a point defect formed via a second order process that is responsible for the observed type inversion effect in oxygen lean material after $\gamma$-irradiation [48, 50, 51]. It was detected so far in three charge states (-, 0 and +), with two levels in the band-gap, a donor level in the lower part ($E_V$ + 0.23 eV) and an acceptor level at the middle of the gap ($E_C$ - 0.55 eV). This center is stable up to temperatures of 325 °C [52]. The corresponding TSC peaks are shown in Fig. 2a. The quadratic dose dependence ($2^{nd}$ order generation) of the $I_p$-center was previously found both for the STFZ and DOFZ diodes up to irradiation doses of 2.80 MGy [48, 50, 51] and is displayed in Fig. 2b up to 5 MGy.

The present TSC investigations, extended up to 5 MGy, indicate a saturation tendency for the $I_p$-center at about $2 \times 10^{12}$ cm$^{-3}$ (STFZ curves in Fig. 2b). The defect structure is still under debate. Due to its generation via a second order process during irradiation at ambient temperature (when mainly single vacancies and interstitials are mobile) it was suggested that the best candidate is the long searched for defect complex $V_2O$. The $V_2O$ is the only one defect generated by a second order process that was evidenced by Electron Paramagnetic Resonance [53, 54] as direct result of irradiation. On the other hand, there are studies based on the annealing of $V_2$ at high temperatures of diodes irradiated with small dose values that revealed the transformation of $V_2$ into another defect $X$ in oxygen rich silicon [55, 56]. In these publications the $X$-defect was associated also with the $V_2O$-complex. However, the energy levels of the $X$ and $I_p$ are very different and cannot both be attributed to the same defect. Contrary to the $I_p$-center no influence of the $X$-defect on the detector performance was observed.

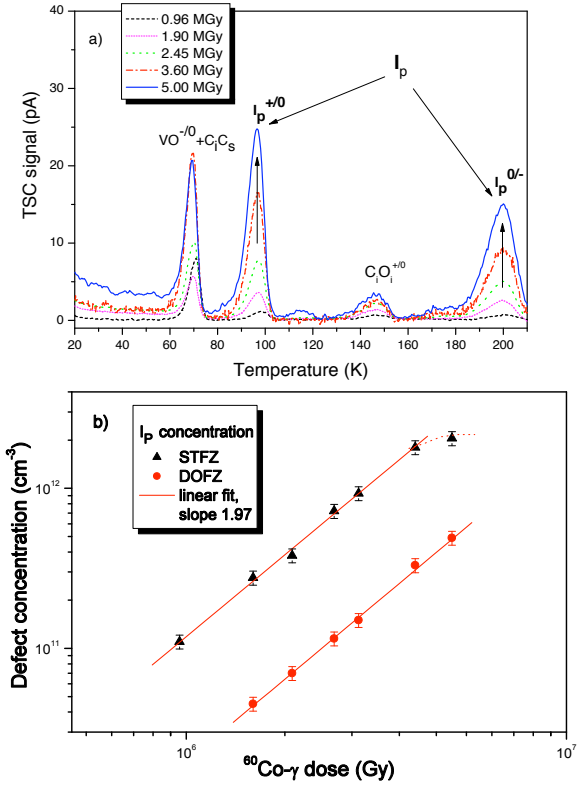

Figure 2: a) TSC spectra recorded on STFZ diodes after γ-doses between 0.96 and 5 MGy; b) γ-dose dependence of the $I_p$-defect concentration, note: saturation seen for STFZ at 5 MGy, slope in log-log scale presentation: 2.0, i.e. quadratic dose dependence.

*4.2. Shallow donor BD*

The $BD$-center is a bistable donor (point defect) strongly generated in oxygen rich material [48, 57]. This center and its bistability have been first observed after $^{60}$Co-γ irradiation in high resistivity DOFZ material. In this case only for one of the defect configurations a clear TSC peak was recorded (see $BD_A$ in Fig. 3a). The other configuration was suggested by the appearance of a large TSC signal in the low temperature range (not shown here). One of the configurations is more stable if the material is exposed to day light ($BD_B$) while the other configuration ($BD_A$) starts to appear when the material is kept some time in the dark at RT (see Fig. 3a). The needed storage at RT in the dark for getting the full transformation from $BD_B$ to $BD_A$ is decreasing with increasing irradiation dose suggesting that, similar to the earlier thermal donors in silicon, the defect configuration depends on the position of the Fermi level. The $BD$-center has a linear dose dependence, indicating that it is generated via a first order process (see Fig. 3b).

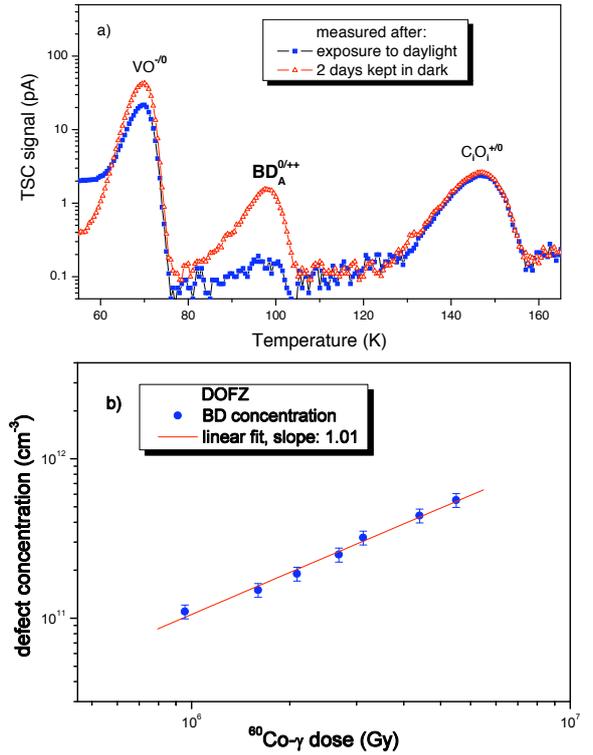

Figure 3: a) TSC spectra corresponding to forward injection of 2 mA for 30 s at 20 K recorded on DOFZ diodes after exposure to day light and after keeping the diode in dark at RT for 2 days; b) γ-dose dependence of the $BD$-defect concentration, slope in log-log scale presentation: 1.0, i.e. linear dose dependence.

A clear evidence for the existence of the two configurations of this defect was obtained in medium doped epitaxial layers irradiated with neutrons [57]. As it can be observed in Fig. 4a, two clear TSC peaks associated with the two defect configurations could be detected. The TSC peaks recorded for different bias voltages have shown that for both defect configurations the emission rate of electrons increases with the electric field. This effect is clearly observed in Figs. 4b and 4c where the TSC peak positions are shifting toward lower temperatures when increasing the applied reverse bias while their integrals are increasing. The analysis of the corresponding TSC peaks revealed that the field enhanced emission rate is due to the Poole-Frenkel effect for both defect configurations, thus indicating, that the defect is a donor in the upper part of the gap. Details about the analysis of TSC signals in the case of Coulomb centers (three-dimensional Poole-Frenkel effect) with accounting for the spatial distribution of the electric field inside the samples can be found in [58]. The zero field activation enthalpy was determined to be $\Delta H_a = E_C - 0.15$ eV for the $BD_B$ configuration (transition between + and ++ charge states) and $\Delta H_a = E_C - 0.24$ eV for $BD_A$ (transition between 0 and ++ charge states). Except for the oxygen lean STFZ material, the $BD$ is detected in all other materials, independent of the type of irradiation. As a point defect it is strongly generated after $\gamma$-irradiation but also detected after neutron and proton damage [34].

The bistability, donor activity and energy levels associate the $BD$-center with the earlier thermal double donors in Si [59–61]. These species of thermal donors can exist in two different structural configurations: one associated with the common shallow double donor state ($TDD$) with occupancy levels $(0/+)$ and $(+/++)$ at $E_C - 0.07$ eV and $E_C - 0.15$ eV, respectively, and a second configuration that forms an Anderson negative U system [62] with a single occupancy level $(0/++)$ at $E_C - 0.22$ eV.

### 4.3. Predictions from the defect analysis for $N_{eff}$ and $I_{dep}$

Both centers, i.e. the $I_p$ as well as the $BD$, have a direct influence on the effective doping concentration. The $I_p$-center contributes with negative space charge while the $BD$-center introduces positive space charge. The contribution of these two centers to $N_{eff}$ has been calculated according to eq. 2 and the results are presented in Fig. 5a together with the values extracted from C-V measurements performed at RT. There is an excellent agreement between measured and calculated values for both STFZ and DOFZ devices. In the case of oxygenated material (DOFZ), the $BD$-defect concentration overcompensates the negative space charge introduced by the $I_p$-center leading to the observed slight increase of the effective doping concentration (Fig. 5a).

Due to its mid-gap acceptor level the $I_p$-center contributes significantly also to the reverse current at RT. The values of $I(V_{dep})$, calculated according to eq. 3, by considering only the $I_p$-center, are shown in Fig. 5b together with the values extracted from I-V measurements performed at RT. Also in this case the predictions deduced from the microscopic measurements represent the macroscopic findings fairly well. Especially the considerably lower reverse current in DOFZ is nicely reproduced and can therefore be attributed to the suppression of the $I_p$-center due to the large O-concentration.

These results represent the first breakthrough in understanding of macroscopic deterioration effects on the basis of a detailed defect analysis and demonstrate that the beneficial oxygen effect in FZ silicon after electromagnetic irradiation results not only in the suppression of deep acceptors (as predicted by the defect models considering the formation of the defect complex $V_2O$ [63]) but also in the creation of a bistable donor similar to the earlier stage thermal double donors $TDD2$ in oxygen rich silicon. These donors can even over-compensate the negative space charge introduced by deep acceptors ($I_p$) such that no type inversion appears in DOFZ material even after very high dose values. Although no identification of the $TDD2$ was accomplished so far, it is well accepted that the oxygen dimers ($O_{2d}$) are part of the defect structure [64].

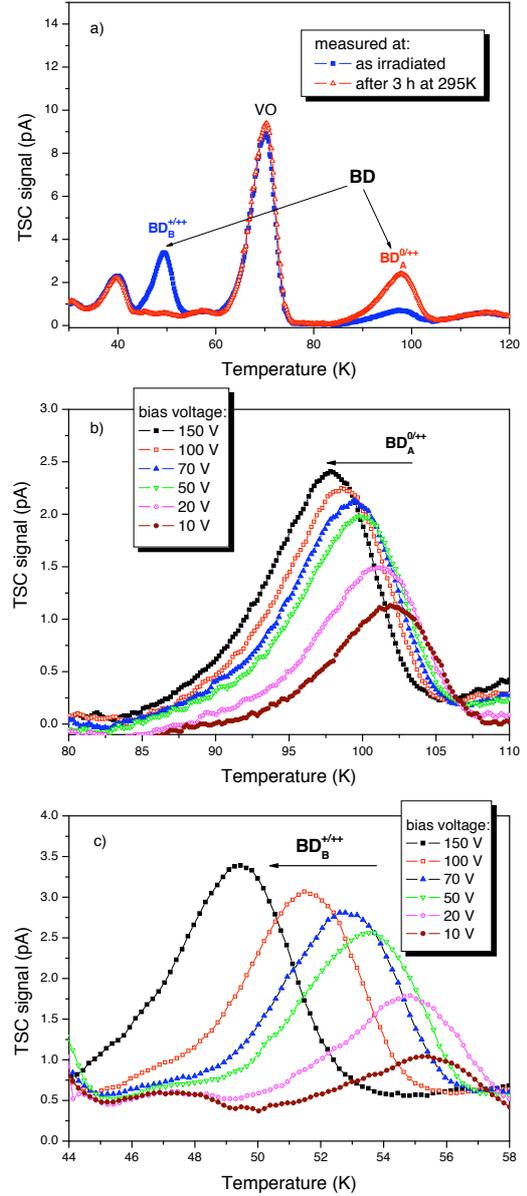

Figure 4: Measurements on EPI-DO diodes after neutron irradiation with $\Phi_{eq} = 5 \times 10^{13}$ cm$^{-2}$; a) TSC spectra corresponding to forward injection of 2 mA for 30 s at 5 K, recorded after exposure to day light and after keeping the diode in dark at RT for 3 h; b) Poole-Frenkel shift of the TSC peak corresponding to the $BD_A$ configuration; c) Shift of the TSC peak corresponding to the $BD_B$ configuration.

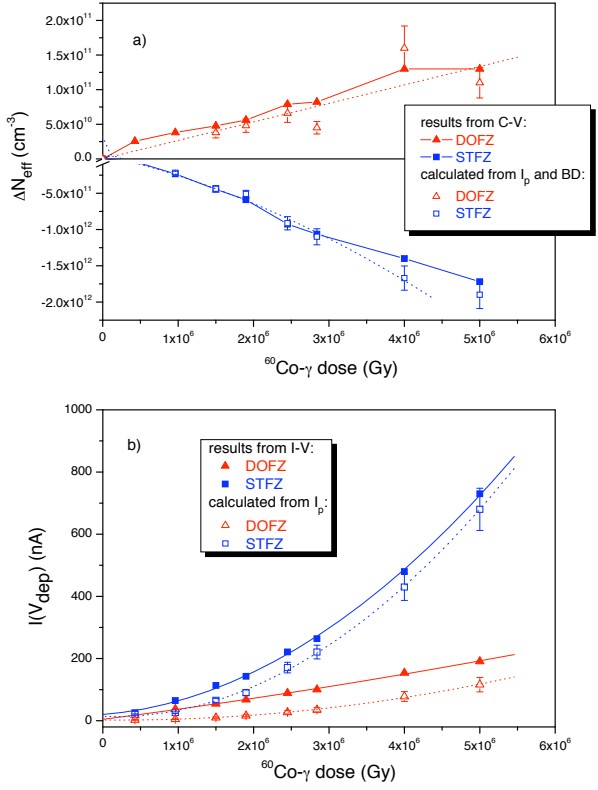

Figure 5: Calculated and measured dose dependence; a) for the irradiation induced effective space charge concentration; b) for the reverse current at full depletion, both for STFZ and DOFZ diodes.

## 5. Defect studies after hadron irradiation

Contrary to low energetic particles, primarily electrons as secondary particles after gamma irradiation (photo- and Compton-effect), producing mainly point defects, more energetic particles, especially hadrons (pions, protons, neutrons or heavy ions) and to some extent also larger energy electrons (see section 6) may lead to a cascade of many successive displacements, resulting in the formation of a large diversity of extended defects as densely packed conglomerates ("clusters") of vacancies and interstitials. The present knowledge on defect clustering is however very limited. Only recently some clear evidences for the formation of electrically active extended defects were reported [58, 65, 66]. Irradiation experiments with reactor neutrons and 23 GeV protons have revealed that there is a group of cluster related defects with direct impact on the device characteristics at operating temperature. In addition a donor, not generated after gamma irradiation, is found here, also visible at moderate energy electron irradiations (see section 6). These defects are discussed below. For this investigation 72 $\mu$m EPI-DO and EPI-ST as well as 280 $\mu$m MCz diodes had been used and annealing studies were performed at 80 °C.

### 5.1. Deep acceptors

The $H(116K)$, $H(140K)$ and $H(152K)$ centers, as already detected earlier [58], are traps for holes with acceptor type levels in the lower part of the band gap that show the Poole-Frenkel effect. As Coulomb centers, their emission rates depend on the local electric field. The parameters for the zero field emission rates reported in Ref. [58] are: $\sigma_p^{116K} = 4 \times 10^{-14}$ cm$^2$ and $\Delta H_a^{116K} = 0.33$eV, $\sigma_p^{140K} = 2.5 \times 10^{-15}$ cm$^2$ and $\Delta H_a^{140K} = 0.36$eV as well as $\sigma_p^{152K} = 2.3 \times 10^{-14}$ cm$^2$ and $\Delta H_a^{152K} = 0.42$eV.

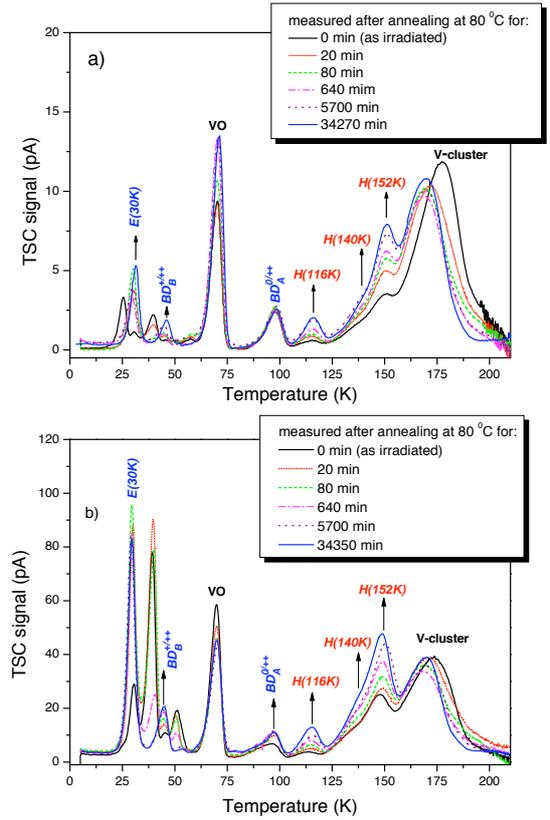

Figure 6: a) TSC spectra recorded on EPI-DO diodes after neutron irradiation with $\Phi_{eq} = 5 \times 10^{13}$ cm$^{-2}$ after different annealing times at 80 °C, measurements after forward injection of 2 mA for 30 s at 5 K; b) same as in a) but after 23 GeV proton irradiation with $\Phi_{eq} = 2.3 \times 10^{14}$ cm$^{-2}$.

These centers were first detected after irradiation with 1 MeV neutrons by means of the TSC technique (see Fig. 6a) but are not seen after gamma irradiation, a strong indication that they are cluster related. We have also detected them after irradiation with 23 GeV protons and the results of TSC investigations are shown in Fig. 6b. As to be seen the defect concentrations are increasing with annealing time. As the defect levels are located in the lower part of the gap, they contribute fully with negative space charge to $N_{eff}$ and are responsible for the long term annealing (so called "reverse annealing", see below). Results after an isochronal annealing stage of 300 °C clearly demonstrate the mentioned Poole-Frenkel shift for these hole traps and in addition a *H(125K)* defect (Fig. 7a). Here it is also demonstrated that the $I_p$-defect does not play a significant role after hadron irradiation.

*5.2. Shallow donors*

The *E(30K)*-defect, also clearly identified in Figs. 6a and 6b, is acting as a trap for electrons. Also this defect was only detected after hadron irradiation (and to some extent after energetic electron irradiation), but not after gamma irradiation. However contrary to the deep hole traps discussed above its generation rate after proton irradiation is at least a factor 6 larger than for neutron damage (compare Fig. 6a and 6b). As for the hole traps discussed above, the *E(30K)*-defect is generated mainly after irradiation, during the first 20 min at 80 °C, a finding subject to further investigation. The enhanced generation after proton irradiation suggests it to be an isolated "point defect" produced via the abundant low energy transfers after Coulomb interaction (further discussion in section 6). The TSC investigations have shown that *E(30K)* is a defect with enhanced-field-emission described by the Poole-Frenkel effect, as best seen after proton irradiation (see Fig. 7b) thus evidencing that it has a donor level in the upper part of the gap [67, 68]. The parameters for the zero field emission rate describing the experimental results are: $\sigma_n^{30K} = 2.3 \times 10^{-14}$ cm$^2$ and $\Delta H_a^{30K} = E_C - 0.1$ eV from the conduction band. This center contributes in its full concentration with positive space charge to $N_{eff}$ and may be consequently partly responsible for the so called "beneficial annealing" effect.

*5.3. Material dependence*

A comparison of TSC spectra obtained in EPI-ST, EPI-DO and MCz diodes after the same neutron irradiation of $5 \times 10^{13}$ cm$^{-2}$ is displayed in Fig. 8 as example for long term annealing at 80 °C. The different thicknesses of the MCz diode (280 $\mu$m) and the EPI ones (72 $\mu$m) have been taken into account for proper normalization. One can clearly see that, while the overall H-center concentration is about the same for all different materials with a slight enhancement of the *H(140K)* in MCz, the $BD_A$ formation is much larger in the EPI-DO than in the EPI-ST diode. For MCz we have only a considerably smaller *BD*-component, an effect which had been observed before and may be due to the comparatively larger O-dimer concentration in EPI-diodes resulting from the growth process [34].

*5.4. Comparison between "microscopic" predictions and "macroscopic" results*

The results from the TSC defect investigations were used to predict the annealing effects of $N_{eff}$ which are then compared with those from C-V measurements at RT. For a complete evaluation of the effective doping concentration at the different annealing stages in addition to the cluster related deep hole centers and the *E(30K)* donor also the contribution from the well known *BD*'*s* are taken into account. Another contribution, which could not be extracted from our TSC measurements, was the removal of the P-dopant donors via formation of the *E*-center (Vacancy - Phosphorus complex). It was added here as a constant contribution not depending on annealing effects. The comparison between the predictions for EPI-DO, EPI-ST and MCz diodes as extracted from the defect analysis and the macroscopic findings are displayed in Figs. 9a, 9b and 9c for neutron and Fig. 10 for proton damage. Especially for the neutron irradiation of the EPI-diodes, where the *E*-center concentration could be reliably extracted macroscopically using the "Hamburg model" (see e.g. [17, 18]), the agreement between prediction and results for the annealing effects is surprisingly good. It should be noted that the MCz diode with its much lower initial doping concentration of only $4.9 \times 10^{12}$ cm$^{-3}$ undergoes type inversion during annealing (around 1000 min at 80 °C), which is very well represented by the microscopic results. Indeed around this annealing time the effective negative space charge given by the net defect concentration equals the initial doping concentration. For the proton irradiation case the large contribution of the $E(30K)$ formation is clearly seen, increasing only during the first 20 minutes at 80 °C and remaining at a constant level thereafter (see Fig. 10).

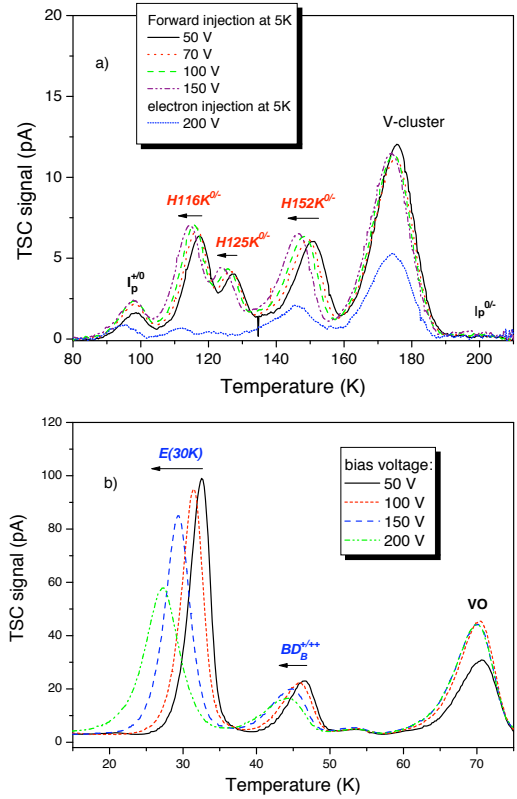

Figure 7: Poole-Frenkel shift of acceptor and donor peaks with electric field, TSC measurements performed on EPI-DO; a) Effect for deep hole traps, neutron irradiation with $\Phi_{eq} = 5 \times 10^{13}$ cm$^{-2}$, isochronal annealing step at 300 °C (see Fig. 10); b) Effect for donor peaks, proton irradiation with $\Phi_{eq} = 2.3 \times 10^{14}$ cm$^{-2}$, measured after isothermal annealing for 11460 min at 80 °C.

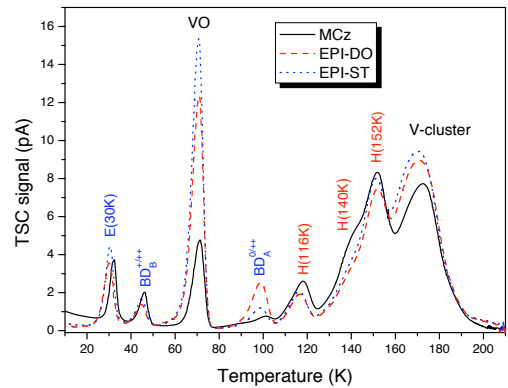

Figure 8: Comparison of TSC spectra measured on MCz, EPI-DO and EPI-ST diodes after neutron irradiation with $\Phi_{eq} = 5 \times 10^{13}$ cm$^{-2}$ after isothermal annealing of 16980 min at 80 °C.

In addition to the isothermal annealing so far discussed, predictions from the defect analysis were also demonstrated for isochronal annealing up to 300 °C. An example is given in Fig. 11a and 11b for a neutron irradiated EPI-DO diode. It is worth noting that the vacancy cluster (containing the $V_2$-center and other probably higher order vacancy complexes) is appreciably narrowing down at higher temperatures, partly due to the annealing of $E4$, $E5$ and $E6$ as also seen in the DLTS-study discussed below. The concentrations of the deep hole centers $H(116K)$ and $H(125K)$ (the latter not visible in the isothermal annealing studies performed at 80 °C) are steadily increasing, while those for $H(140K)$ and $H(152K)$ start to anneal out after reaching a maximum for an annealing temperature around 200 °C. Details of this temperature dependence may be of considerable value for understanding the nature of these defects. The $E(30K)$ donor concentration reaches a maximum after annealing at 240 °C and drops down rapidly at higher annealing temperatures. This and the steady increase of the $BD_B$ strength, transforming into the normal thermal donor, will also have an impact above 200 °C. Finally the concentration of the *E*-center (donor removal), annealing out above 150 °C, could not be measured in this study as well as other defect transformations (e.g. of divacancy in the *X*-center for temperatures above 200 °C). However, despite all these uncertainties and disregarding the influence of the donors, the main temperature dependence for $\Delta N_{eff}$ (from C-V) is determined by the total concentration of the deep hole traps, as displayed in Fig.11b. Denoted as acceptors here are only the deep hole traps $H(116K)$, $H(125K)$, $H(140K)$ and $H(152K)$ and as donors, the $E(30K)$, $BD$ and $TDD$.

### 5.5. Striking difference between neutron and proton irradiation

The most obvious difference between neutron and proton irradiation is given by the introduction rate of donors. While the $BD$ generation rates are quite similar for neutron resp. proton induced damage ($1.6 \times 10^{-2}$ cm$^{-1}$ resp. $2 \times 10^{-2}$ cm$^{-1}$), those for the $E(30K)$-center are largely different ($9 \times 10^{-3}$ cm$^{-1}$ for

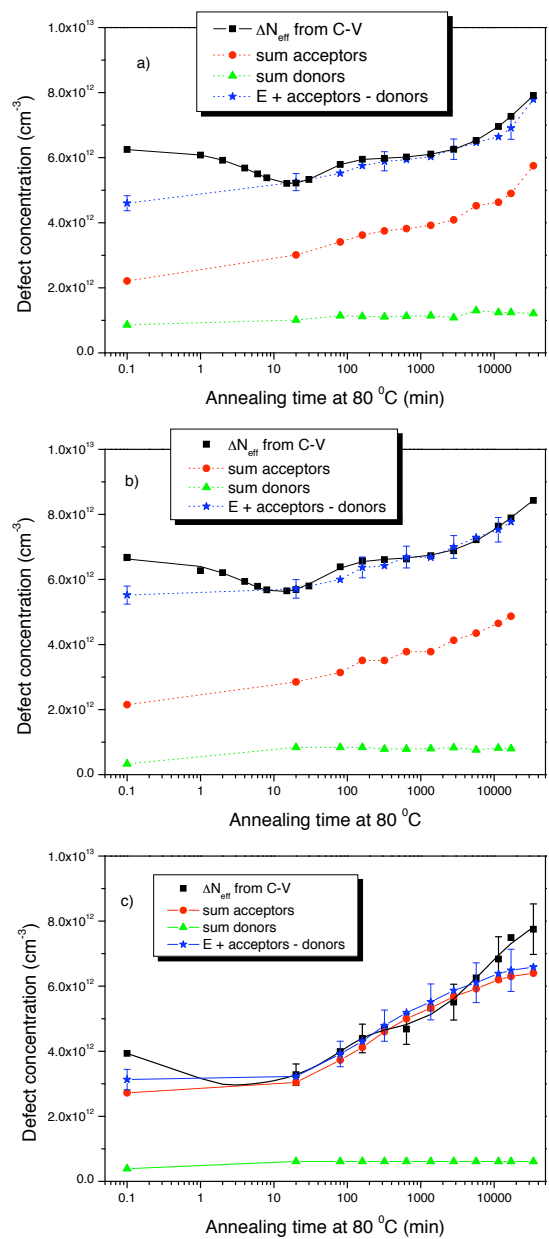

Figure 9: Comparison of the measured and calculated change of the irradiation induced effective space charge concentration after neutron irradiation with $\Phi_{eq} = 5 \times 10^{13}$ cm$^{-2}$ for 80 °C annealing (see text); a) EPI-DO; b) EPI-ST; c) MCz.

n- versus $6 \times 10^{-2}$ cm$^{-1}$ for p-irradiation). For short annealing times around the minimum of $\Delta N_{eff}$ (so called "stable damage component" in the Hamburg model) the large $E(30K)$ concentration after proton irradiation leads to an over-compensation of the deep hole centers and thus, in contrast to neutron damage, the effective doping remains positive, increasing with fluence (Fig. 12).

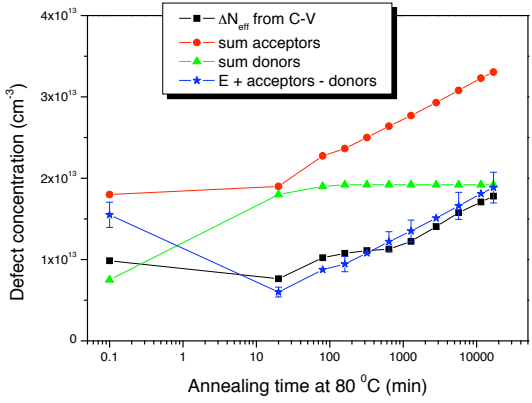

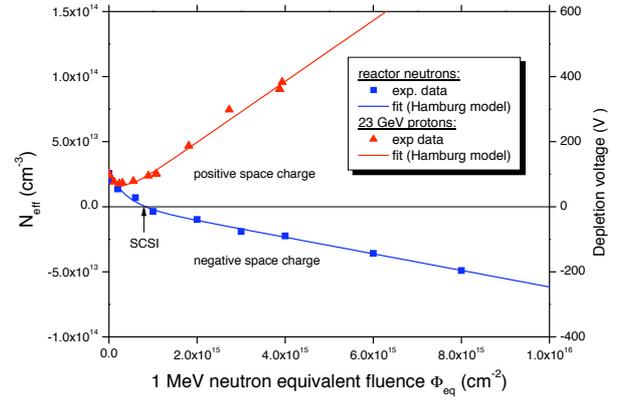

Figure 10: Same as in Fig. 9 but for an EPI-DO diode irradiated by 23 GeV protons with $\Phi_{eq} = 2.3 \times 10^{14}$ cm$^{-2}$.

Figure 12: Effective space charge concentration, extracted from C-V measurements in EPI-DO diodes as function of fluence for reactor neutron and 23 GeV proton irradiation.

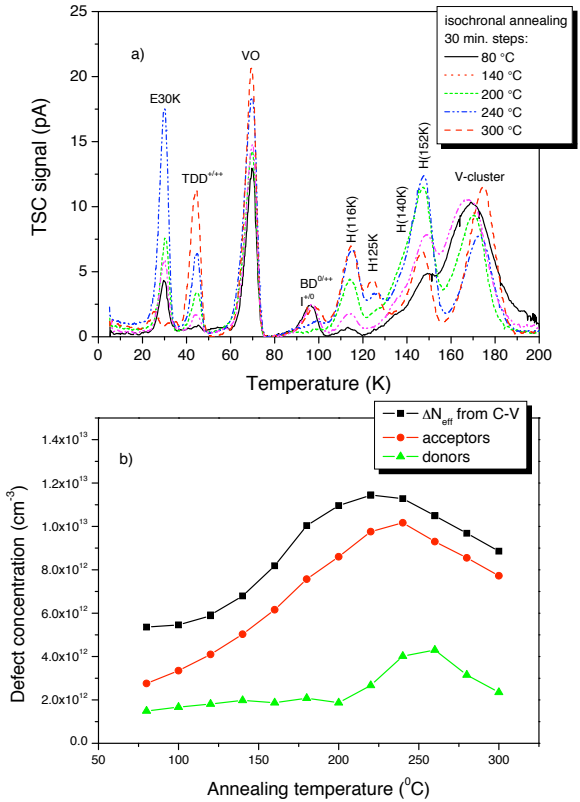

Figure 11: Measurements taken at isochronal annealing between 80 and 300 °C on an EPI-DO diode, after neutron irradiation with $\Phi_{eq} = 5 \times 10^{13}$ cm$^{-2}$; a) examples of TSC spectra at several temperature steps; b) change of effective space charge concentration, extracted from C-V curves, in comparison to defect concentrations extracted from TSC results.

### 5.6. Correlation with reverse current

As mentioned in section 2 the reverse current is much more sensitive already to small fluence levels and thus related defects can be well studied with DLTS measurements. For hadron irradiation the responsible defects are likely cluster dominated and thus we concentrated our measurements on mid gap defect levels. Contrary to the isothermal annealing studies of $N_{eff}$ at 80 °C described before, we present here results of isochronal annealing experiments performed with a neutron irradiated MCz diode ($\Phi_{eq} = 3 \times 10^{11}$ cm$^{-2}$) in the temperature range between 20 °C and 200 °C. One example of the obtained DLTS spectra before and after annealing at different temperatures is shown in Fig. 13a. The broad DLTS peak centered at around 200 K is composed of at least 4 cluster related electron traps which are labeled $E4$, $E5$, $E6$ and $V_2^{-/0}$. While the levels $E4$ and $E5$ [65, 69–71] decrease even at moderate temperatures and vanish after annealing at 100 °C for 30 minutes, the defect complex $E6$ (referred to as $E205a$ in reference [72]) is more stable and anneals out in the temperature range between 100 °C and 200 °C. The concentrations of the cluster related defects can not be obtained directly from the DLTS spectra, due to their overlap with the singly charged state of the di-vacancy. Therefore, difference spectra during the annealing process were analyzed, i.e. the difference between the spectrum before annealing and those recorded after each annealing step. Fig. 13b shows the change of the reverse current versus the corresponding change of the defect concentration, taken during the isochronal annealing in 20 °C steps with an annealing time of 30 minutes each. The defects which were taken into account are the levels $E5$ and $E6$. Also the change of the reverse current refers to the difference between the value before annealing and those obtained after each annealing step. It is clearly seen, that the reverse current and the cluster related defect concentration are correlated. The dashed line indicates a possible linear correlation.

The formation of all these cluster related defects is not affected by the oxygen content of the material or the silicon growth process suggesting that they are complexes of multi-vacancies and/or multi-interstitials located inside extended disordered regions. Presently there are no decisive experiments which would clarify the identity of the observed cluster related defects $H(116K)$, $H(140K)$, $H(152K)$, $E(30K)$, $E4$, $E5$ and $E6$. Correlated fundamental studies with structure sensitive methods (e.g. Infrared Absorption and EPR) have to be performed in the future.

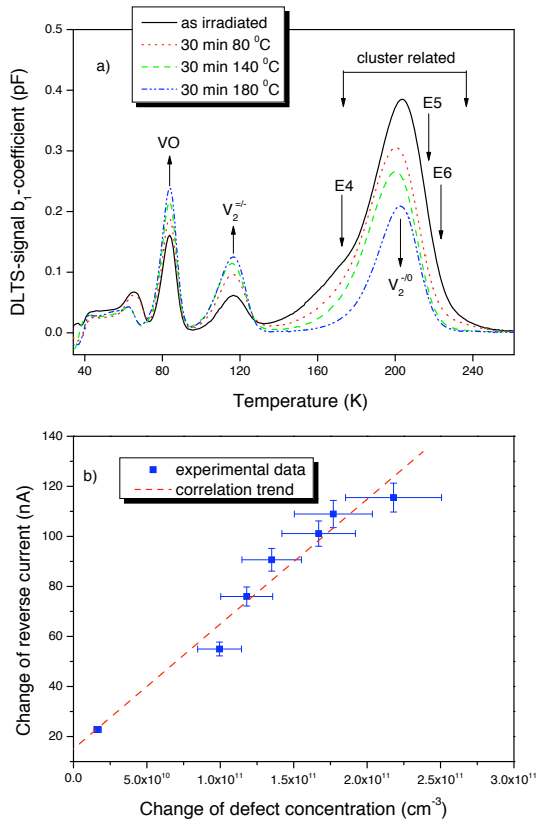

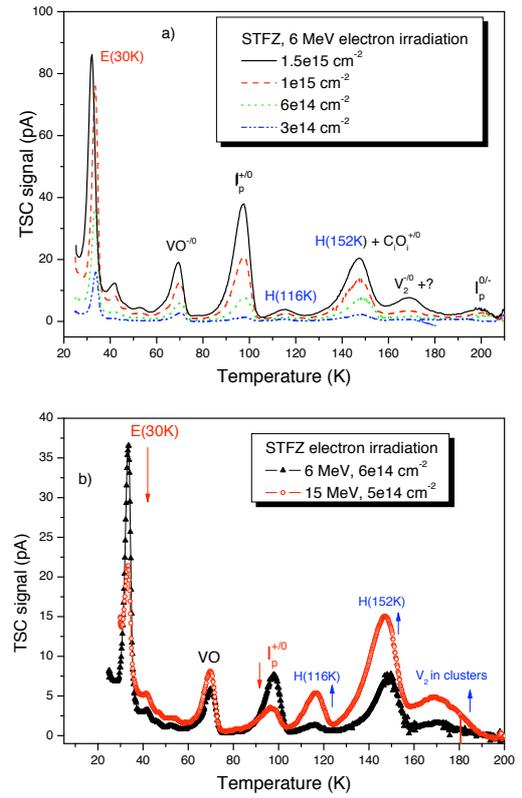

Figure 13: a) DLTS spectra measured on a MCz diode after neutron irradiation with $\Phi_{eq} = 3 \times 10^{11}$ cm$^{-2}$, examples of isochronal annealing steps between 20 and 200 °C; b) correlation between the changes of defect concentration as measured by DLTS and the reverse current at full depletion as extracted from I-V characteristics.

Figure 14: a) TSC spectra measured on STFZ diodes irradiated by 6 MeV electrons with 4 different fluences, measurements performed with forward injection at 25 K before annealing; b) comparison of TSC spectra as measured in STFZ diodes irradiated by 6 and 15 MeV electrons.

## 6. Studies after electron irradiation

As discussed in section 2., irradiations with several MeV electrons should already result in the generation of cluster related defects as seen after hadron irradiations but depending on the energy the ratio between cluster and point defects will vary. In order to test this assumption we reinvestigated some TSC measurements after 6 and 15 MeV electron irradiation. The spectra as seen directly after irradiation (no annealing performed in this case) are shown in Fig. 14a (6 MeV electron irradiation with 4 different fluences) and Fig. 14b (comparison between 6 and 15 MeV).

The spectra are very similar to those observed after hadron irradiations however the following remarks should be taken into account. The peak labeled $E(30K)$ is surely the same donor defect as seen in neutron and much more pronounced in proton irradiation (see Fig. 6a and 6b). It is therefore most likely produced by small PKA energies as expected from Coulomb interaction of charged hadrons or electrons. It is however not seen after $\gamma$ irradiation and could tentatively be attributed to the tri-interstitial (minimum threshold energy 75eV) as observed in photoluminescence measurements [46]. For the electron irradiations only oxygen lean float zone diodes (STFZ) had been used and therefore the generation of the $BD$-defect, as visible in the hadron irradiated oxygen rich EPI-diodes is very unlikely. Instead at 98 K a large peak, rapidly growing with increasing fluence is detected and identified as the +/0 level of the $I_p$-defect (compare with Fig 2a). Also the 0/- level of $I_p$ (at about 200 K) is visible but possibly due to the implications by the deep cluster defects the occupation probability is reduced. At the large temperature range the deep hole traps as discovered in hadron irradiations are clearly visible here too. Comparing the spectra for 6 and 15 MeV electron irradiation (Fig. 14b) it is clearly seen that the concentration of the deep hole traps ($H(116K)$, $H(140K)$, $H(152K)$) is increasing with increasing energy thus supporting the expectation that the cluster formation should largely increase from 6 to 15 MeV. On the other hand the $E(30K)$ concentration is even slightly decreasing, hence again supporting the conclusion that it is not cluster related but an isolated point defect. Finally in Fig. 15 the extracted defect concentrations for 6 MeV irradiations are plotted as function of electron fluence. The concentrations for $E(30K)$ and the deep hole traps are proportional to the fluence whereas for the 98K defect the fit function is doubtless quadratic (compare with Fig. 2b after $\gamma$-irradiation), confirming again that the $I_p$-defect, i.e. the $V_2O$-center, is generated via a second order process.

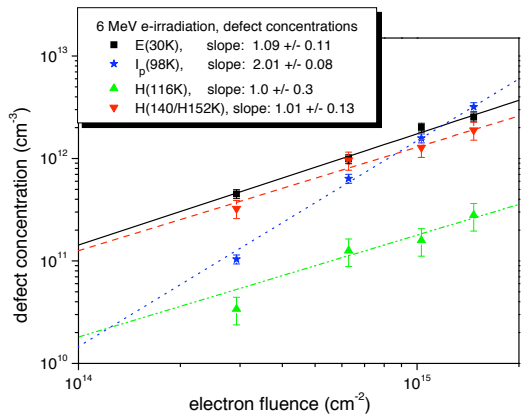

Figure 15: Dependence of defect concentrations on 6 MeV electron fluence, representation in log-log scale, dotted lines show fits through data, slope of 1 indicates linear dependence for $E(30K)$ and the deep hole traps $H(116K)$, $H(140K)$ + $H(152K)$, while the $I_p$-defect at 98 K shows a purely quadratic dependence as expected (see Fig. 2b).

## 7. Summary

Future colliding beam experiments at e.g. the S-LHC with hadron fluences up to several $10^{16}$ cm$^{-2}$ in the innermost layers of the tracking area will place an unprecedented challenge for the radiation tolerance of silicon sensors not yet met by present day devices. Improvements have shown to be possible by certain modifications of the silicon material, process technology and operational conditions. But success oriented endeavors need to be based on a detailed understanding of the damage effects in the silicon bulk and their implications on detector performance. For the first time a close correlation between the findings by defect spectroscopy and the resulting damage effects in the detector properties are demonstrated in a comprehensive study. A first breakthrough in this respect was achieved by detailed investigations of the damage effects after intense $\gamma$-irradiation, where only point defect generation is possible. Both the change of the space charge concentration (depletion voltage) and the increase of the reverse current are now completely understood in terms of radiation induced defects.
Experiments have been extended up to 5 MGy using standard and oxygen enriched FZ diodes. While the deep acceptor $I_p$ with a level close to mid-gap controls both the increase of negative space charge as well as the generation current, the shallow bistable donor $BD$ is responsible for the increase of positive space charge. The introduction rate of both defects depends on the oxygen content of the material. Especially the quadratic dose dependence of the $I_p$-defect concentration, indicating a second order process, the strong reduction of its introduction rate in oxygen rich material and the thermal stability have lead to a possible assignment with the long searched for $V_2O$-defect center. In oxygen enriched material the generation of the $BD$ prevails and over-compensates the $I_p$-center such that the diodes stay at positive space charge up to 5 MGy.

In recent years it was then also possible to obtain very promising results for the much more complicated case of hadron induced damage. Experiments have been performed using standard and oxygen enriched EPI as well as MCz diodes. It is shown, that in addition to point defects a diversity of extended defects as densely packed conglomerates (clusters) of vacancies or interstitials play a dominant role leading to the observed change in the space charge concentration and hence depletion voltage. The three defect centers $H(116K)$, $H(140K)$ and $H(152K)$ are hole traps and show the well known Poole-Frenkel effect, demonstrating that they are acceptors with levels in the lower half of the band gap. Hence they contribute fully with negative space charge to the effective doping concentration at room temperature. On the other hand two shallow donors ($E(30K)$ and $BD$) are positively charged at room temperature and therefore partly compensate the negative space charge induced by the acceptors. In contrast to the deep acceptors the $E(30K)$ donor generation after proton irradiation is strongly enhanced with respect to neutron damage, thus explaining that for proton irradiation the effective space charge remains positive while after neutron damage type inversion is observed. The $E(30K)$-defect was only detected after hadron and electron irradiation and is likely related to multi-interstitials. In contrast to $\gamma$-irradiation the formation of the $BD$ and $I_p$-defects play a comparatively minor role in hadron irradiation.

Finally we like to emphasize that in addition to previous investigations for $\gamma$-irradiation the damage effects after hadron irradiation, especially the annealing behavior, as seen at "macroscopic scale" can now be understood by the "microscopically" investigated kinetics of the responsible defects. The cluster related defects detected so far and presented in this work seem to be independent on the material. These results toward a complete understanding are very promising. They are regarded to be an excellent basis for further dedicated defect engineering. The present investigations will be extended to larger fluences and including a systematic comparison between neutron and proton damage. An improvement of the radiation hardness is possible only by compensation of the cluster related acceptors with enhanced donor generation, as achieved in oxygen enriched material, or by inhibiting the vacancy cluster effects as likely possible via hydrogen enrichment. Especially the latter possibility will be pursued by us in the future.

## 8. Acknowledgements

This work was carried out in the frame of the CERN-RD50 collaboration and funded partly by the CiS Hamburg project under Contract No. SSD 0517/03/05, the BMBF under Contract No. 05HC6GU1 and by the Romanian Ministry of Education and Research under the Core Program, contract 45N/2009. We gratefully thank E. Nossarzewska at ITME, Warsaw, for her professional work in preparing the EPI-diodes and performing the spreading resistance measurements. Likewise the help of A. Barcz at ITE, Warsaw, was invaluable in doing the SIMS measurements. All diode processing had been done by CiS, Erfurt, managed by R. Röder, and we like to express our many thanks for the very successful cooperation. Many thanks are also especially due to G. Kramberger for the neutron irradiation at the TRIGA reactor at JSI Ljubljana, Z. Li for carrying out the $\gamma$-irradiation at BNL and M. Glaser for the proton irradiation at the CERN PS facility and B.G. Svensson for providing the electron irradiations at Stockholm. One of us, I. Pintilie, is deeply indebted to R. Klanner, University of Hamburg, for support during several guest stays and numerous discussions. She also likes especially to thank the Humboldt foundation for providing a Humboldt fellowship to her for conducting main parts of this work.